\documentclass[11pt,twoside]{article}

\usepackage{asp2006}
\usepackage{epsf}
\usepackage{psfig}
\usepackage{lscape}

\begin{document}
\title{Testing the Evolutionary Link Between SMGs and QSOs: are Submm-Detected QSOs at $z\sim2$ `Transition Objects' Between These Two Phases?}   
\author{Kristen Coppin}   
\affil{Institute for Computational Cosmology, Durham University, Durham, DH1 3LE, UK}    

\begin{abstract} 
Local spheroids show a relation between their masses and those of the
super-massive black holes (SMBH) at their centres, indicating a link
between the major phases of spheroid growth and nuclear
accretion. These phases may correspond to high-z submillimetre galaxies (SMGs)
and QSOs, separate populations with surprisingly similar redshift
distributions which may both be phases in the life cycle of individual
galaxies, with SMGs evolving into QSOs.  Here we briefly discuss our recent results in \citet{Coppin08}, where we have tested this connection by weighing the black holes and mapping CO in submm-detected QSOs, which may be transition objects between the two phases, and comparing their baryonic, dynamical and H$\alpha$-derived SMBH masses to those of SMGs at the same epoch.  Our results split our sample of submm-detected QSOs into two categories (although a bigger sample would probably show a continuous trend): (1) CO is detected in 5/6 very optically luminous ($M_\mathrm{B}\sim -28$) submm-detected QSOs with BH masses $M_\mathrm{BH}\simeq10^9$--10$^{10}$\,M$_\odot$, confirming the presence of large gas reservoirs of $M_\mathrm{gas}\simeq3.4\times10^{10}$\,M$_\odot$.  Our BH masses and dynamical mass constraints on the host spheroids suggest, at face value, that these optically luminous QSOs at $z=2$ lie about an order of magnitude above the local BH-spheroid
 relation, $M_\mathrm{BH}/M_\mathrm{sph}$. However, we find that their BH masses are $\sim30$ times too large and their surface density is
 $\sim300$ times too small to be related to typical SMGs in an
 evolutionary sequence.  (2) We measure weaker CO emission in
 four fainter ($M_\mathrm{B}\sim-25$) submm-detected QSOs with properties, BH masses ($M_\mathrm{BH}\simeq5\times10^8$\,M$_\odot$), and
 surface densities similar to SMGs.  These QSOs appear to lie near the local
 $M_\mathrm{BH}/M_\mathrm{sph}$ relation, making them plausible
 `transition objects' in the proposed evolutionary sequence linking
 QSOs to the formation of massive young galaxies and BHs at high-$z$.

\end{abstract}


\vspace{-1cm}
\section{Introduction}   

It appears that every massive, local
spheroid hosts a SMBH in its centre whose mass is proportional to that
of its host (e.g.\ \citealt{Magorrian98}; \citealt{Gebhardt00}). This
suggests that the black holes (BHs) and their surrounding galaxies
were formed synchronously.  This hypothesis has found support from
hydrodynamical simulations of galaxy formation, which use feedback from
winds and outflows from active galactic nuclei (AGN) to link the growth
of the SMBH to that of its host (e.g.\ \citealt{DiMatteo05}; 
\citealt{Hopkins05}; \citealt{Bower06}). 
Thus these models support a picture, first presented by
\citet{Sanders88}, where a starburst-dominated ultra-luminous
infrared galaxy (ULIRG), arising from a merger, evolves first into an
obscured QSO and then into an unobscured QSO, before finally becoming a
passive spheroid.

The high-redshift population of ULIRGs in this proposed evolutionary cycle are the 
submillimetre (submm) galaxies (SMGs; \citealt{SIB97}).  These systems have
ULIRG-like bolometric luminosities, $L_\mathrm{IR}\geq 10^{12}$\,L$_\odot$, and they have many of the properties expected for gas-rich mergers
(e.g.\ \citealt{Tacconi06}).  This population
evolves rapidly out to a peak at $z\sim 2.3$, crudely matching
the evolution of QSOs  and providing
additional circumstantial evidence for a link between SMBH growth and
spheroids \citep{Chapman05}.  Two further results have shed light on the evolutionary
link between SMGs and QSOs:  (1) A modest fraction of optically luminous QSOs at $z\sim 2$ are
detected in the submm/mm ($\sim25$\%; \citealt{Omont03}) showing that
the QSO- and SMG-phases do not overlap significantly, given the lifetime estimates of the two populations 
(QSOs make up $\sim 4$\% of flux-limited samples of SMGs; \citealt{Chapman05}).  But
when a QSO {\it is} detected in the submm/mm then it is likely to be in the 
transition phase from an SMG to an unobscured QSO, making its
properties a powerful probe of the evolutionary cycle 
(e.g.\ \citealt{Stevens05});  (2) The evolutionary state of the SMBHs within SMGs can also be judged using the 2-Ms {\it Chandra} Deep Field North observations to derive accurate AGN luminosities and hence lower limits on the BH masses ($M_\mathrm{BH}$) in those SMGs with precise redshifts in this region (\citealt{Alexander05nat,Alexander05}; \citealt{Borys05}).  These studies suggest that the AGN in typical SMGs are growing almost continuously -- but that the SMBHs in these galaxies appear to be several times less massive than seen in comparably massive galaxies at $z\sim 0$ (\citealt{Alexander07}; see also D.M. Alexander's paper in these conference proceedings). Together these results argue for a fast transition from a star-formation-dominated SMG-phase to the AGN-dominated QSO-phase \citep{Page04}. Subsequent rapid black hole growth is then required to account for the present-day relation between spheroid and SMBH masses. {\it Can we confirm this and more generally test the proposed evolutionary link between SMGs and QSOs at the peak of their activity
at $z\sim 2$?}

Here we briefly discuss the results of \citet{Coppin08}, who carried out a quantitative test of the proposed link between $z\sim2$ SMGs and QSOs.  We have obtained precise systemic redshifts from UKIRT near-infrared spectroscopy of potential transition QSOs (i.e.~submm/mm-detected QSOs) spanning $1.7<z<2.6$ and then used the IRAM Plateau de Bure Interferometer (PdBI) to search for CO emission.  We relate their dynamical, gas and SMBH masses to SMGs from the PdBI CO survey of \citet{Greve05} and explore the the proposed evolutionary sequence which links QSOs to the formation of massive young galaxies and SMBHs at high redshift.

\vspace{-0.5cm}
\section{Comparison of the Submm-Detected QSOs and SMGs}

\subsection{Gas Content and Star-Formation Efficiencies}

We detect CO emission from six of the ten submm-detected
QSOs in our sample, confirming that they contain a significant amount of molecular gas and that a large fraction of the mm emission is from starbursts.  
Adopting a CO-to-gas conversion factor appropriate for local galaxy populations exhibiting similar
levels of star formation activity to submm-bright galaxies or QSOs (e.g.\ ULIRGs; \citealt{Solomon05}), we derive a median gas mass of the entire sample (i.e. including non-detections) of $(2.5\pm0.7)\times 10^{10}$\,M$_{\odot}$, similar to that found for $z\sim 2$ SMGs and for $z>4$ QSOs.
  
The star formation efficiency (SFE) is a measure of how effective a galaxy is at converting its gas into stars and can be represented by the continuum-to-line ratio, or $L_\mathrm{FIR}/L'_\mathrm{CO}$, as this ratio presumably traces the star formation rate per total amount of gas in a galaxy. Our submm-detected QSOs have a median of $L_\mathrm{FIR}=(8.0\pm1.9)\times10^{12}$\,L$_\odot$, and a star-formation rate (SFR) for our sample following \citet{Kennicutt98} of SFR=$1360\pm320$\,M$_\odot\,\mathrm{yr}^{-1}$. The star formation efficiencies of our QSOs are also comparable to those measured for SMGs, $250\pm100$\,L$_\odot\,\mathrm{(K\,km\,s^{-1}\,pc^{2})^{-1}}$, suggesting that the gross properties of the star formation in the QSOs are like those seen in SMGs. We determine that the submm-detected QSOs have enough cold gas to sustain
their current episode of star formation for $\tau_\mathrm{depletion}\sim \mathrm{M_{gas}/SFR}\sim2.5\times10^{10}$\,M$_{\odot}/1360$\,M$_\odot\,\mathrm{yr}^{-1}\sim20$\,Myr, implying a submm-detected QSO phase lifetime of $2\times\tau_\mathrm{depletion}=40$\,Myr.

\subsection{Dynamical Masses}

CO line widths can be directly converted into dynamical masses, assuming a size and inclination for the gas reservoirs. We note that we find a lower incidence of bimodal CO line profiles in the QSOs, compared to SMGs, which we believe results from a selection bias towards lower average inclination angles for the QSOs.  The median CO line width of our sample is $550\pm180\,\mathrm{km\,s^{-1}}$.  Adopting a 2\,kpc scale size for the gas distribution in the QSOs and a typical inclination of 20$^\circ$ (assuming that the submm-detected QSOs are seen at lower inclination angles, see e.g.~\citealt{Alexander07}) we derive a median dynamical mass of M$(<2$\,kpc$)\sim(2.1\pm1.4)\times 10^{11}$\,M$_\odot$, similar to SMGs (which are assumed to be randomly orientated, $i$=30$^\circ$).   

\subsection{BH masses and Evolutionary Status}

\begin{figure}
\psfig{file=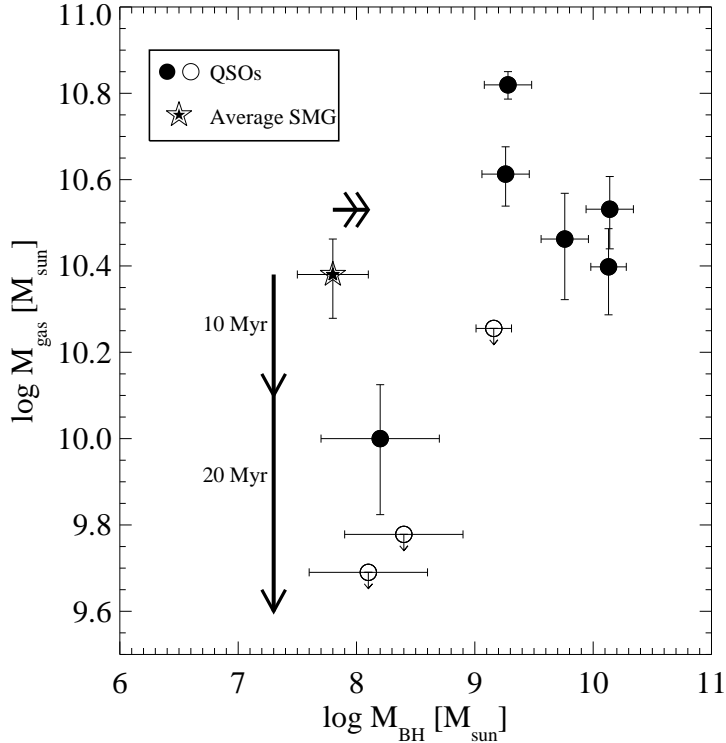,width=0.8\textwidth}
\caption{$M_\mathrm{gas}$ as a function of $M_\mathrm{BH}$ for our
sample of CO-observed submm-detected QSOs.  The arrows (offset slightly for clarity)
indicate the movement of an average SMG in terms of its gas mass
depletion and synchronous BH growth in arbitrary scalable timecales of
10 and 20\,Myr, assuming a SFR of 1000\,M$_\odot$\,yr$^{-1}$ and
Eddington-limited BH growth.  This demonstrates that an SMG could
evolve into $M_\mathrm{BH}\simeq10^{8}$\,M$_\odot$,
$M_\mathrm{gas}\simeq0.6\times10^{10}$\,M$_\odot$ submm-detected QSOs
in a reasonable timescale, whereas an average SMG would need
substantially ($\sim10\times$) more gas and more time
($>\mathrm{SMG\,\,lifetime}$) to evolve into the
$M_\mathrm{BH}\simeq4\times10^{9}$\,M$_\odot$,
$M_\mathrm{gas}\simeq3.0\times10^{10}$\,M$_\odot$ submm-detected QSOs.
This suggests that the submm-detected QSOs hosting BHs of
$M_\mathrm{BH}\simeq10^{8}$\,M$_\odot$ with $M_\mathrm{gas}<1\times10^{10}$\,M$_\odot$ comprise `transition objects' that we can
use to probe the intermediary evolutionary stage between the SMG and
luminous QSO phases, while the rarer more luminous QSOs with
$M_\mathrm{BH}> 10^{9}$\,M$_\odot$ are not related to typical
SMGs.}
\label{fig:dave}
\end{figure}

We determine $M_\mathrm{BH}$ for our QSOs using the \citet{Greene05} virial $M_\mathrm{BH}$ estimator which calculates the BH mass from the observed H$\alpha$ or H$\beta$ emission line widths and fluxes. We find a median black hole mass in our submm-detected QSO sample of $(1.8\pm1.3)\times10^{9}$\,M$_\odot$, which is about an order of magnitude larger than SMGs \citep{Alexander07}. Combined with our dynamical estimates of the spheroid mass above, these yield $M_\mathrm{BH}/M_\mathrm{sph}\sim 9\times10^{-3}$.  This $M_\mathrm{BH}/M_\mathrm{sph}$ ratio for this sample of submm-detected QSOs at $z=2$ is an order of magnitude larger than the local ratio of \citet{Haring04}, suggesting that BH growth occurs more rapidly than the bulge formation in $z\sim2$--3 submm-detected QSOs, although $M_\mathrm{sph}$ suffers from large uncertainties due to the unknown CO radii and inclination angles.  This ratio is also significantly above that seen for SMGs at $z\sim 2$ (\citealt{Alexander07}; see also D.M. Alexander's paper in these proceedings).  However, this comparison masks a broad range in BH masses within our submm-detected QSO sample and so we split the sample into two subsets based on their BH masses.

Looking at the optically luminous submm-detected QSOs in
our sample we find that we detect CO emission in 5/6 of them.
However, the estimated BH masses for these QSOs, $M_\mathrm{BH}\simeq10^{9}$--10$^{10}$\,M$_{\odot}$, are too large (and their number densities too small) for them to be related to typical SMGs in a simple evolutionary cycle.  We propose that the progenitors of these most massive QSOs are a rare subset of SMGs with
$M_\mathrm{gas}>4\times10^{11}$\,M$_\odot$ with a number density of
$\simeq10\,\mathrm{deg}^{-2}$ which will be possible to detect with
future SCUBA-2 surveys.

For the optically less luminous ($\sim L^{\star}$)
submm-detected QSOs, we marginally detect one source in CO and obtain sensitive
limits for three further QSOs.  The BH masses for these systems are
$M_\mathrm{BH}\simeq10^{8}$\,M$_\odot$, similar to the estimates for BHs
in SMGs.  These submm-detected QSOs are consistent with being `transition' objects between SMGs and submm-undetected QSOs, as it is feasible to
link their BH masses to those of SMGs by Eddington limited growth
for a period comparable to the gas depletion timescale of the QSOs,
$\sim 10$\,Myrs.  The space density of these QSOs is also in rough
agreement with that expected for the descendents of SMGs given current
estimates of the relative lifetimes of QSOs and SMGs.  We conclude that
these $\sim L^{\star}$, $M_\mathrm{BH}>10^{8}$\,M$_\odot$ submm-detected QSOs are consistent with being in a very brief prodigious star
formation phase, and that they simply do not possess sufficiently large gas
reservoirs to sustain the SFR (which is why these might be less often
detected in CO), although a larger sample of CO observations of
submm-detected QSOs with these BH masses is required for confirmation.

\vspace{-0.5cm}
\section{Final Remarks}

To make further progress on understanding the evolutionary
links between SMGs and QSOs requires a larger survey of the submm and
CO emission from typical QSOs ($M_\mathrm{B}\approx -25$), which could be undertaken in the near future with LABOCA and SCUBA-2.  In addition, measurements of other CO
transitions for the submm-detected QSOs (e.g.~from IRAM 30-m, ALMA, 
EVLA, and SKA) are required to place better constraints on the temperature and density of the
molecular gas and thus provide a more accurate determination of the
line luminosity ratios and hence total gas masses of these
systems. Similarly, higher resolution CO observations are essential to
put strong constraints on the reservoir sizes and inclination angles, and hence $M_\mathrm{dyn}$, 
needed to compare the two populations.  Finally, better measurements of
the far-infrared SEDs (with SABOCA, SCUBA-2 or \textit{Herschel}) will
yield more accurate measures of $L_\mathrm{FIR}$ and $T_\mathrm{dust}$
for submm-detected QSOs to constrain the contribution from an AGN component.

\acknowledgements 
I thank the UK Science and Technologies Facilities Council for a postdoctoral fellowship and the following collaborators for allowing me to present this research: A.M. Swinbank, R. Neri, P. Cox,  D.M. Alexander, Ian Smail, M.J. Page, J.A. Stevens, K.K. Knudsen, R.J. Ivison, A. Beelen, F. Bertoldi, A. Omont.



\begin{thebibliography}{}
\bibitem[Alexander et al.(2005a)]{Alexander05nat}Alexander D.M. et al., 2005a, Nat., 434, 738
\bibitem[Alexander et al.(2005b)]{Alexander05}Alexander D.M. et al., 2005b, ApJ, 632, 736 
\bibitem[Alexander et al.(2008)]{Alexander07}Alexander D.M. et al., 2008, AJ, 135, 1968
\bibitem[Borys et al.(2005)]{Borys05} Borys C. et al., 2005, ApJ, 635, 853
\bibitem[Bower et al.(2006)]{Bower06} Bower R.G. et al., 2006, MNRAS, 370, 645
\bibitem[Chapman et al.(2005)]{Chapman05}Chapman S.C. et al., 2005, ApJ, 622, 772
\bibitem[Coppin et al.(2008)]{Coppin08}Coppin K., et al., 2008, MNRAS, 389, 45
\bibitem[Di~Matteo, Springel \& Hernquist(2005)]{DiMatteo05}Di~Matteo T., Springel V. \& Hernquist L., 2005, Nat., 433, 604
\bibitem[Gebhardt et al.(2000)]{Gebhardt00}Gebhardt K. et al., 2000, ApJ, 539, 13
\bibitem[Greene \& Ho(2005)]{Greene05} Greene J.E. \& Ho L.C., 2005, ApJ, 630, 122
\bibitem[Greve et al.(2005)]{Greve05} Greve T.R. et al., 2005, MNRAS, 359, 1165
\bibitem[H\"{a}ring \& Rix(2004)]{Haring04}H\"{a}ring N. \& Rix H.-W., 2004, ApJ, 604, L89
\bibitem[Hopkins et al.(2005)]{Hopkins05}Hopkins P.F. et al., 2005, ApJ, 630, 705
\bibitem[Kennicutt(1998)]{Kennicutt98}Kennicutt R.C., 1998, ARA\&A, 36, 189
\bibitem[Magorrian et al.(1998)]{Magorrian98}Magorrian J. et al., 1998, AJ, 115, 2285
\bibitem[Omont et al.(2003)]{Omont03}Omont A. et al., 2003, A\&A, 398, 857
\bibitem[Page et al.(2004)]{Page04}Page M.J. et al., 2004, ApJ, 611, L85
\bibitem[Sanders et al.(1988)]{Sanders88}Sanders D.B. et al., 1988, ApJ, 328, L35
\bibitem[Smail, Ivison \& Blain(1997)]{SIB97} Smail I., Ivison R.J. \& Blain A.W., 1997, ApJL, 490, L5
\bibitem[Solomon \& Vanden Bout(2005)]{Solomon05} Solomon P.M. \& Vanden Bout P.A., 2005, ARA\&A, 43, 677
\bibitem[Stevens et al.(2005)]{Stevens05}Stevens J.A. et al., 2005, MNRAS, 360, 610
\bibitem[Tacconi et al.(2006)]{Tacconi06}Tacconi L.J. et al., 2006, ApJ, 640, 228
\end{thebibliography}
\end{document}